# Ba$_9$RE$_2$(SiO$_4$)$_6$ (RE=Ho-Yb): A New Family of Rare-earth based Honeycomb Lattice Magnets

Andi Liu[1], Fangyuan Song[1], Zhaohu Li[1], Malik Ashtar[1], Yuqi Qin[1], Dingjun Liu[1], Zhengcai Xia[1], Jingxin Li[2], Zhitao Zhang[2], Wei Tong[2], Hanjie Guo[3#], Zhaoming Tian[1,4]*

1 Wuhan National High Magnetic Field Center and School of Physics, Huazhong University of Science and Technology, Wuhan, 430074, China

2 Anhui Province Key Laboratory of Condensed Matter Physics at Extreme Conditions, High Magnetic Field Laboratory, HFIPS, Chinese Academy of Sciences, Hefei, Anhui 230031, China

3 Songshan Lake Mat Lab, Neutron Sci Platform, Dongguan 523808, China

4 Shenzhen Huazhong University of Science and Technology Research Institute, Shenzhen, Guangdong 518057, China.

**ABSTRACT:** Rare-earth (RE) based honeycomb-lattice materials with strong spin-orbit coupled $J_{eff}=1/2$ moments have attracted great interest as a platform to realize Kitaev quantum spin liquid (QSL) state. Herein, we report the discovery of a new family of RE based honeycomb-lattice magnets Ba$_9$RE$_2$(SiO$_4$)$_6$ (RE=Ho-Yb), which crystallize into the rhombohedral structure with space group $R\bar{3}$. In these serial compounds, magnetic RE$^{3+}$ ions are arranged on a perfect honeycomb lattice within the *ab*-plane and stacked in the "ABCABC"-type fashion along the *c*-axis. All Ba$_9$RE$_2$(SiO$_4$)$_6$ (RE=Ho-Yb) polycrystals exhibit the dominant antiferromagnetic interaction and absence of magnetic order down to 2 K. In combination with the magnetization and electron spin resonance (ESR) results, distinct anisotropic magnetic behaviors are proposed for compounds with different RE ions. Moreover, the synthesized Ba$_9$Yb$_2$(SiO$_4$)$_6$ single crystals show large magnetic frustration with frustration index $f=\theta_{CW}/T_N>8$ and no long-range magnetic ordering down to 0.15 K, being a possible QSL candidate state. These serial compounds are attractive for exploring the exotic magnetic phases of Kitaev materials with *4f* electrons.

■ **INTRODUCTION**

Quantum spin-liquid (QSL) is an exotic magnetic phase hosting long-range spin entanglement and fractional spinon excitation,[1-3] in which the interacting spins remain disordered at zero temperature due to the strong quantum fluctuations. Since the initial proposal of resonance-valence bond (RVB) state in triangular-lattice spin systems by Anderson in 1973,[4] the physical nature and experimental identification of QSL state have attracted great interest in condensed matter physics. In a long-time duration, the explorations on QSL state have mainly associated with the geometrically frustrated magnets (GFMs), because the geometric frustration can enhance the spin competing interactions and quantum fluctuations. A large variety of GFMs with different Archimedean lattices have been studied to search for the QSL states,[1,5-7] including two-dimensional triangular-lattice and kagome-lattice antiferromagnets as well as three-dimensional pyrochlore-lattice magnets. Recently, another theoretical model on realizing the QSL state was proposed by Alexei Kitaev in the honeycomb-lattice spin-1/2 systems driven by the bond-dependent Ising spin-exchange interactions,[8] usually called Kitaev-QSL. In 2009, George Jackeli and Giniyat Khaliullin proposed a mechanism for designing anisotropic Kitaev spin interactions through spin-orbital coupling (SOC),[9] subsequently it stimulated the exploration of Kitaev-QSL state in honeycomb-lattice materials containing heavy-element-based magnetic ions due to their intrinsic large SOC interactions.[10-13] Additionally, beyond the importance of fundamental physics, recent studies have also proposed that the Kitaev-QSL can host topological order and non-Abelian statistics,[14,15] the later property is crucial for application in quantum computation. Therefore, the experimental realizations of Kitaev-QSL have both fundamental and practical importance.

The search for Kitaev QSL state in real materials have been mainly performed on the honeycomb-lattice magnets comprising the 4d/5d transition-metal ions and *4f* Rare-earth (RE) ions.[2,6,10] Compared to the 4d/5d electrons, the *4f* electrons can exhibit much stronger SOC and more localized magnetic exchange interactions, thus can yield highly anisotropic interactions between effective $J_{eff}$=1/2 state.[16-18] Another difference is that the RE-based materials exhibit antiferromagnetic (AFM) Kitaev interactions instead of the ferromagnetic (FM) ones of the 4d/5d systems, that allow them to realize the zero magnetic-field QSL ground state as well as field-tuned QSL phase.[19-21] Indeed, intensive studies have recently been conducted on a number of RE-based Kitaev candidates with various *4f* electron configurations including trichlorides $YbX_3$ (*X* = F, Cl, Br and I),[22-24] $A_2REO_3$ (A = Li and Na; RE=Pr,Tb) [20,25,26] and $RE_2Si_2O_7$ (RE=Er-Yb) oxides,[27-29] and chalcohalides RE*Ch*X (Ch = O, S, Se, and Te).[17,30] Unfortunately, the experimental identification on Kitaev QSL states remain to be a challenge, most of the discovered RE-based Kitaev materials have classical magnetic ordered phase at low temperatures. For examples, the relatively well studied $YbCl_3$ exhibits a short-range magnetic order below 1.20 K and long-range AFM order with Néel temperature $T_N$~ 0.60 K,[22] $Na_2PrO_3$ enter into the AFM state below $T_N$~ 4.6 K.[25] Moreover, from the structure viewpoint, magnetic RE ions in $YbCl_3$, $Yb_2Si_2O_7$ and $Na_2PrO_3$ compounds are arranged on the distorted honeycomb-lattice not on the perfect honeycomb-lattice,[22,26,27] which can introduce extra weak asymmetric exchange interactions precluding the formation of Kitaev-QSL State. Thus, engineering of Kitaev interaction in new materials with perfect honeycomb lattice are imperative to study the Kitaev-QSL physics.

In this work, we present the synthesis and magnetic properties on a new family of rare-earth $Ba_9RE_2(SiO_4)_6$ (RE=Ho-Yb) orthosilicate, which crystallize into the rhombohedral structure with space group R$\bar{3}$. For these serial compounds, several family members have been synthesized and characterized for their excellent thermal stability and luminescent property.[31-33] While, the lattice topology of magnetic RE ions and their magnetism have not yet been reported, motivating the present study. The structural analysis reveals that magnetic $RE^{3+}$ ions are located on an ideal honeycomb-lattice in the *ab* plane and then the honeycomb layers are stacked in the "ABCABC"-type fashion along the c-axis. Based on the magnetic characterization and electron spin resonance (ESR) results, we unveiled the dominant antiferromagnetic (AFM) interactions and magnetic anisotropy of $Ba_9RE_2(SiO_4)_6$ (RE=Ho-Yb). Moreover, the synthesized $Ba_9Yb_2(SiO_4)_6$ single crystal shows the AFM interaction with $\theta_{CW}$ ~1.2 K and no long-range magnetic order down to 0.15 K, being a possible Kitaev-QSL candidate.

## ■ EXPERIMENTAL SECTION

**Material synthesis.** The series of $Ba_9RE_2(SiO_4)_6$ (RE=Ho-Yb) polycrystalline samples were prepared by the conventional solid-state reaction method. The stoichiometric $BaCO_3$(99.99%), $SiO_2$ (99.99%) and RE oxides (99.9%) were used as starting materials, which were mixed and thoroughly ground for 3 hours. Then, the pre-powders were sintered in the temperature range from 1000°C-1300 °C for totally 4 days with several intermediate grindings to get the polycrystalline samples.

The $Ba_9Yb_2(SiO_4)_6$ single crystals were grown by a high-temperature flux method, similar to the previous reports.[34,35] The polycrystalline powders were placed in 40 mL platinum crucibles and heated in an electric furnace at 1300 °C for 24 hours to ensure the homogeneous melting. Then, the furnace was cooled down slowly at a rate of 3-5 °C/h to 800°C. After cooling down to room temperature, the $Ba_9Yb_2(SiO_4)_6$ single crystals were mechanically separated.

**Structure Characterization.** The polycrystalline samples of $Ba_9RE_2(SiO_4)_6$ were characterized by the room temperature powder X-ray diffraction with Cu Kα radiation ($\lambda$ = 1.5418 Å) in the range of $2\theta$ = 10-80°. Rietveld refinements on the Powder XRD patterns were performed using the general structure analysis system (GSAS). For $Ba_9Yb_2(SiO_4)_6$ single crystal, the crystal structure was determined by single-crystal X-ray diffraction (SXRD) with the diffraction data collected by a Kappa Apex2 CCD diffractometer (Bruker) using graphite-monochromated Mo Kα radiation ($\lambda$ = 0.71073 Å). Subsequently, the structure was refined using the ShelXL least-squares refinement.

**Physical property measurements:** Magnetic measurements were carried out using a superconducting quantum interference device (SQUID, Quantum Design) in applied DC magnetic field ($\mu_0H$) up to 7 T, including the temperature dependent magnetic susceptibility from 1.8 K-300 K and isothermal field-dependent magnetization $M(\mu_0H)$ measurements. The pulsed field magnetizations up to 42 T were measured by the induction method at Wuhan National High Magnetic Field Centre (WHMFC) with a calibration by the DC magnetization data. The X-band (9.4 GHz) electron spin resonance (ESR) measurements were carried out using a Bruker spectrometer at the High Magnetic Field Laboratory of the Chinese Academy of Science. The specific heat ($C_p$) of $Ba_9Yb_2(SiO_4)_6$ single crystal was measured on physical property measurement system (PPMS) equipped with a $^3$He cooling stage down to 0.15 K.

## ■ RESULTS AND DISCUSSION

**Structural description**

All synthesized $Ba_9RE_2(SiO_4)_6$ (RE=Ho-Yb) samples are isostructural and crystallized into the rhombohedral structure with space group $R\bar{3}$ (No. 148), similar to the previous report of $Ba_9Sc_2(SiO_4)_6$.[32] The schematic crystal structure of $Ba_9Yb_2(SiO_4)_6$ as a representative is presented in Figure 1a. The structure framework is mainly constructed by corner communion of $SiO_4$-$YbO_6$-$SiO_4$ layers, the $YbO_6$ octahedra are linked by two $SiO_4$ tetrahedral units in each layer. In the unit cell, there are nine crystallographic sites, Yb atoms occupy a single location on site 6c coordinated by six oxygen ions forming the distorted $YbO_6$ octahedra. The Si atoms are located on site 18f and Ba atoms have three sites on 3a, 6c, 18f, respectively. The detailed coordination environments of Yb, Ba1, Ba2, Ba3 and Si cations with surrounding oxygen ions are depicted in Figure 1b. The related interatomic distances and bond angles are listed in Table 1. From that, the distorted $YbO_6$ octahedra have three shorter Yb-O bonds of 2.1804(3) Å and three longer bonds of 2.2422(3) Å. The O-Yb-O bond angles vary from 89.881(8) to 95.085(8) °. The distorted $SiO_4$ tetrahedra have four different Si-O bond distances ranging between 1.5803(3) Å and 1.66210(18) Å. There are three independent Ba sites coordinated with 12, 9, and 10 oxygen atoms, respectively. The Ba1 occupies a special position with a 3-fold inversion symmetry, which is coordinated by 12 oxygen atoms forming a distorted cuboctahedron. The Ba2 occupies a special position with symmetry 3 and is coordinated to 9 oxygen atoms. Ba3 is surrounded by 10-oxygen ions, where the coordination sphere originates from that of Ba1, half a cuboctahedron is capped by oxygen accompanied with a hexagonal pyramid at the bottom.

In the crystal structure of $Ba_9Yb_2(SiO_4)_6$, the framework of magnetic lattices is constructed by the $YbO_6$ octahedra arranged in a hexagonal array, which is further linked by two $SiO_4$ tetrahedral units. In the *ab* plane, magnetic $Yb^{3+}$ ions are located on a honeycomb lattice with an equal nearest $Yb^{3+}$-$Yb^{3+}$ intralayer distance of $a/\sqrt{3}$ ~ 5.7718 Å (see Figure 1d). Then, the magnetic $Yb^{3+}$ honeycomb layers form the "ABCABC"-type stacking fashion along the *c*-axis with equivalent interlayer distance $c/3$~7.3696 Å (see Figure 1c). Moreover, the magnetic honeycomb layers are separated by several nonmagnetic layers formed by Ba/Si-O polyhedrons, namely, the magnetic $Yb^{3+}$ layers and nonmagnetic $Ba^{2+}$/$Si^{4+}$ cations are alternatively arranged along the *c*-axis.

Figure 2a shows the experimental and refined XRD patterns of $Ba_9RE_2(SiO_4)_6$ (RE=Ho-Yb) powders, the crystal structure of $Ba_9Sc_2(SiO_4)_6$ is used as a starting model for refinement.[35] As seen, the calculated XRD patterns match well with the experimental data, and the refined lattice parameters decrease monotonically with the reduced $RE^{3+}$ ionic radius as shown in Figure 2b. For detailed structural parameters, the bond distances, bond angles, the intraplane and interplane RE−RE distances of the four compounds are summarized in Table 1 and Table S1 (see Supporting Information). For the as-grown $Ba_9Yb_2(SiO_4)_6$ single crystals, a typical photograph with the dimensions of 2.0 × 3.0 × 0.2 mm³ is shown in Figure 2c, the refined crystallographic data are listed in Table 2. Based on the structure analysis, no site-occupancy disorder is detected between the magnetic $Yb^{3+}$ ions and nonmagnetic $Ba^{2+}$ and $Si^{4+}$ cations in $Ba_9Yb_2(SiO_4)_6$, this is reasonable considering the large difference of ionic radii and coordination numbers between $Yb^{3+}$ and $Ba^{2+}/Si^{4+}$ cations. Then, the $Yb^{3+}$ and $Ba^{2+}/Si^{4+}$ cations prefer to form the fully ordered arrangements without the antisite mixing disorder. This is quite attractive to investigate the intrinsic-ordered honeycomb lattice physics, avoid the puzzle on understanding its magnetic ground state induced by magnetic exchange randomness as reported in the quantum spin liquid (QSL) material $YbMgGaO_4$.[6,18] Similar disorder-free occupancy has been reported in the kagome-lattice $RE_3BWO_9$ and triangular-lattice $NaYbCh_2$ (Ch=O, S, and Se).[36,37]

**Magnetic property of $Ba_9RE_2(SiO_4)_6$ polycrystals**

Temperature dependence of magnetic susceptibilities $\chi$(T) for $Ba_9RE_2(SiO_4)_6$ (RE=Ho-Yb) polycrystalline samples were measured from 2 to 300 K in an applied magnetic field $\mu_0H$ = 0.5 T. The inverse susceptibilities $1/\chi$(T) were fitted by the Curie-Weiss (CW) law, $\chi = C/(T-\theta_{CW})$, where $\chi$ is the susceptibility, $C$ is the Curie constant and $\theta_{CW}$ is the Curie-Weiss temperature. The effective magnetic moments $\mu_{eff}$ were obtained by the following relationship: $\mu_{eff} = (3k_BC/N_A)^{1/2}$, where $k_B$ is the Boltzmann constant and $N_A$ is the Avogadro's number. Considering that the temperature-dependent magnetization is significantly influenced by crystal electric field (CEF) splitting,[38,39] the CW fits were performed at high temperature paramagnetic (PM) regime of 120-300 K and low-temperature regions of 5-15 K for RE=Er, Yb and 40-70 K for RE=Ho, Tm, respectively. The obtained $\theta_{CW}$ and $\mu_{eff}$ are summarized in Table 3, and the effective moments ($\mu_{fi}$) of free $RE^{3+}$ ions calculated by $g_J[J(J+1)]^{1/2}$ are also provided for comparisons. The isothermal field-dependent magnetizations $M$ ($\mu_0H$) were measured at selected temperatures. Additionally, the X-band electron spin resonance (ESR) were measured to help to understand the magnetic behaviors, the corresponding g factors are calculated as $g = h\nu/\mu_BH_r$, where $h$ is the Planck constant, $\nu$ = 9.4 GHz is the microwave frequency, $H_r$ is the resonance absorption field and $\mu_B$ is the Bohr magneton. The derivative ESR spectra (d$P$/d$H$) ($P$ is the integral ESR intensity) for $Ba_9RE_2(SiO_4)_6$ (RE=Ho-Yb) at different temperatures are depicted below.

The *4f*-$RE^{3+}$ ions can be divided into two categories, Kramers ions with odd 4f electrons and non-Kramers ions with even number of 4f electrons. For the RE-based magnets containing Kramers ions, the magnetic ground state can be described by $J_{eff}$ =1/2 state protected by time-reversal symmetry, like the present investigated $Ba_9RE_2(SiO_4)_6$ (RE=Er, Yb) compounds. While for the materials with non-Kramers ions, usually the crystal symmetry cannot provide enough symmetry operations protecting the degeneracy of the crystal-field levels, the low-lying singlet states will form and be applicable to the $Ba_9RE_2(SiO_4)_6$ (RE=Ho, Tm) systems. From this viewpoint, magnetic study on the $Ba_9RE_2(SiO_4)_6$ (RE=Ho-Yb) samples can provide the a direct comparison on the honeycomb-lattice system.

**$Ba_9Ho_2(SiO_4)_6$.** The magnetic susceptibility $\chi$(T) of $Ba_9Ho_2(SiO_4)_6$ under $\mu_0H$ = 0.5 T is shown in Figure 3a, no long-range magnetic ordering or spin freezing is detected down to 2 K. The high-temperature CW fits to the inverse susceptibility $\chi^{-1}$(T) give rise to $\theta_{CW}$ = -8.89 K and $\mu_{eff}$ = 10.90$\mu_B$/$Ho^{3+}$. The effective moment is close to the value $g_J[J(J+1)]^{1/2}$ =10.6$\mu_B$ for free $Ho^{3+}$ ($^5I_8$) ions with $m_J$ = ±8 doublet ground state. The low-temperature fitted $\theta_{CW}$ = -3.5 K indicates the dominant AFM interactions between $Ho^{3+}$ moments.

Figure 3b presents the isothermal magnetization $M$ ($\mu_0H$) at 2 K, which displays a nonlinear response without saturation up to field of 7 T. From the high-field magnetization curves [see the inset of Figure 3b], the maximum magnetization at 2 K reaches $9.63\mu_B/Ho^{3+}$ in accordance with the value $M_S = g_JJ\mu_B = 10\mu_B/Ho^{3+}$ for the free $Ho^{3+}$ ion with $J=8$. As an effective probe to obtain $g$-factors, the measured X-band ESR spectra at selected temperatures were presented in in Figure 3c. From that, two sets of resonance lines with resonance field at $\mu_0H_{r1}= 0.088$ T and $\mu_0H_{r2}= 0.35$ T, which give two $g$-factors with $g_1$=7.5 and $g_2$=1.9 at 2 K (see Figure 3d). The presence of two resonance fields can be attributed to the low symmetry of distorted $HoO_6$ octahedra and its local coordination environment since the g-factors are dependent on the CEF.[40] While, using the above obtained $g$-factors, the calculated maximum value of saturated magnetization is $gJ_{eff}\mu_B$=$3.75\mu_B$ in case of $J_{eff}$=1/2 ground state, this value is much smaller than the experimental result. This discrepancy suggests the effective $J_{eff}$ =1/2 state cannot be defined to describe the magnetic behaviors at 2 K, the experimental determination on the $J$=8 multiplet splitting of $Ho^{3+}$ ions is necessary by the inelastic neutron scattering, as experimentally performed on the triangular-lattice AFMs $PrZnAl_{11}O_{19}$ with non-Kramers $Pr^{3+}$ ions.[41]

**$Ba_9Er_2(SiO_4)_6$.** The susceptibility $\chi(T)$ of $Ba_9Er_2(SiO_4)_6$ is presented in Figure 4a. The high temperature CW fitting yields $\theta_{CW}$ = -7.53 K and $\mu_{eff}$ = $9.93\mu_B/Er^{3+}$, and the low temperature CW fitting gets $\theta_{CW}$ = -2.43 K and $\mu_{eff}$ = 9.14 $\mu_B/Er^{3+}$. These obtained effective moments are consistent with the expected value $9.59\mu_B$ of free $Er^{3+}$ ions. The negative value of $\theta_{CW}$ indicates the dominant AFM interactions between $Er^{3+}$ moments. Figure 4b shows the $M$ ($\mu_0H$) of $Ba_9Er_2(SiO_4)_6$ at selected temperatures. After subtracting the linear field-dependent magnetization contribution, the saturated magnetization at 2 K reaches $4.85\mu_B$. From the ESR spectra shown in Figure 4c, two sets of resonance lines can be found at temperature below 20 K, one peak is located at ~300 Oe and another one appears at ~1300 Oe. The later one gives $g_2$=4.98 at 2 K, which show the nonmonotonic temperature dependences with minimum value at 15 K [see Figure 4d]. This reflects the variation of surrounding environments of $Er^{3+}$ ions. For the low-field resonance, it can be attributed to the weak exchange interactions of $Er^{3+}$ moments along the easy magnetization direction.

**$Ba_9Tm_2(SiO_4)_6$.** Figure 5a presents the magnetic susceptibility $\chi(T)$ of $Ba_9Tm_2(SiO_4)_6$, which shows no magnetic transitions down to 2 K. The high temperature CW fitting yields $\mu_{eff}$ = $7.90\mu_B/Tm^{3+}$ close to the expected value $7.57\mu_B$ for free $Tm^{3+}$ ion. As decreased temperatures, the $\chi(T)$ exhibits an upturn behavior below 50 K possibly due to the contribution of Van Vleck paramagnetism. This is consistent with the isothermal $M$ ($\mu_0H$) curves showing nearly linear field dependence up to 7 T and no signature of saturation presented in Figure 5b. Moreover, the maximum magnetization $4.6\mu_B$ at 42 T is much smaller than the expected saturated magnetization $M_S = g_JJ\mu_B = 7\mu_B/Tm^{3+}$ for free $Tm^{3+}$ ions. These results further indicate the formation of low-lying singlet states instead of $J_{eff}$=1/2 doublet state. The ESR spectra exhibit two peaks with very weak intensity as shown in Figure 5c, which can be from the isolated paramagentic $Tm^{3+}$ ions. Two values of g-factor with $g_1 \sim 2.56$ and $g_2 \sim 1.99$ at 2 K are obtained, as provided in Figure 5d.

**$Ba_9Yb_2(SiO_4)_6$.** For polycrystalline $Ba_9Yb_2(SiO_4)_6$ sample, no signature of magnetic order is detected down to 2 K from the $\chi(T)$ curves shown in Figure 6a. The high temperature CW fitting yields $\mu_{eff} = 4.86\mu_B/Yb^{3+}$ close to the moment of $4.54~\mu_B/Yb^{3+}$ for free $Yb^{3+}$ ions ($4f^{13}$, $^2F_{7/2}$). As temperature decreased, the $\chi^{-1}(T)$ gradually deviates from the linear dependence with slope change below temperature ~ 100 K reflecting the influence of the CEF effect where more population of electrons will occupy at the CEF ground states at lower temperatures. This effect can also explain the reduced moment $\mu_{eff} = 2.62~\mu_B/Yb^{3+}$ from the low temperature CW fits. Similar $\chi(T)$ behaviors have also been observed in other $Yb^{3+}$-based quantum magnets, such as $Yb(BaBO_3)_3$[35] and $KBaYb(BO_3)_2$[42] compounds. The low temperature fitted negative CW temperature $\theta_{CW} = -1.18$ K suggests the dominant AFM interactions between $Yb^{3+}$ moments.

The $M(\mu_0H)$ curves at selected temperatures of $Ba_9Yb_2(SiO_4)_6$ are shown in Figure 6b. At 2 K, the saturated magnetization reaches ~$1.28~\mu_B/Yb^{3+}$ about the half of the effective magnetic moment. From the ESR spectra of $Ba_9Yb_2(SiO_4)_6$ shown in Figure 6c, two broad resonance lines can be identified at temperatures below 60 K, the resonance peaks are located at field regions between 0.2 - 0.35 T. By using two sets of Dysonian function in Lorentzian form,[43,44] the ESR spectra at 2 K are fitted, as presented in Figure 6d. From that, the g-factor with $g_1 = 2.79$ and $g_2 = 2.34$ are obtained. Based on these estimated g-factors, the calculated $M_{Sat} = g_{ave}J_{eff}\mu_B$ ~ $1.25~\mu_B$ by using $g_{ave} = \sqrt{(g_1^2 + 2g_2^2)/3}$ matches well with the experimental value of polycrystals.

**Magnetic property and specific heat of $Ba_9Yb_2(SiO_4)_6$ single crystal**

In these serial compounds, considering that $Ba_9Yb_2(SiO_4)_6$ is one promising material to realize the Kitaev QSL state with $J_{eff}=1/2$ local moment, the $Ba_9Yb_2(SiO_4)_6$ single crystals were successfully grown to investigate the magnetic ground state and its anisotropy. Figure 7a presents the temperature-dependent magnetic susceptibility under $\mu_0H = 0.5$ T along the c-axis ($\chi_c$, H// c) and the ab-plane ($\chi_{ab}$, H // ab), respectively. In combination with the susceptibilities under field from 0.1 T to 2 T shown in the inset of Figure 7b and Figure S2, no magnetic transition or spin freezing behaviors are detected down to 1.8 K. No magnetic anomaly at ~2.3 K is detected excluding the formation of $Yb_2O_3$ impurity.[45] From the low-temperatures (5-15 K) Curie-Weiss fits, we obtain the effective moments $\mu_{eff,c} = 2.63~\mu_B/Yb^{3+}$ and $\mu_{eff,ab} = 2.44~\mu_B/Yb^{3+}$ and $\theta_{CW,c} = -1.19$ K and $\theta_{CW,ab} = -1.15$ K for H // c and H // ab, respectively. These parameters agree with its counterpart of the polycrystals, the negative CW temperatures along both directions support the dominant AFM interactions between $Yb^{3+}$ moments. In Figure 7b, we present the $M(\mu_0H)$ curves at 2 K. After subtracting a linear-dependent magnetization contribution as denoted by the dashed lines, the saturated magnetization ~ $1.35~\mu_B$ (H // c) and ~ $1.18~\mu_B$ (H // ab) are obtained, these moment values agree well with the g-factor from the ESR spectra with $g_1 = 2.79$ and $g_2 = 2.34$ based on the calculation of $M_s=g_JJ_{eff}\mu_B$ with $J_{eff}= 1/2$. Therefore, $Ba_9Yb_2(SiO_4)_6$ can be considered as an effective $J_{eff}=1/2$ system.

To check the magnetic ground state, zero-field specific heat $C_p(T)$ was measured from 150 K to 0.15 K as presented in Figure 8b. As cooling temperatures, the $C_p(T)$ results reach the minimum value near 2 K then increase steeply down to the lowest accessible temperature, similar to the observation in the Yb-based triangular-lattice systems $Ba_3Yb(BO_3)_3$[46] and $ABaYb(BO_3)_2$ (A=Rb,K).[42,47] The absence of magnetic anomaly in the $C_p(T)$ curves suggest the high quality of single crystals. By integrating the magnetic contribution to specific heat, the change of magnetic entropy in temperature region 0.15 K- 5K is far from the value of $R \ln 2$, indicating that most entropy remains below 0.15 K. Here, the large magnetic frustration index $f = |\theta_{CW}/T_N| > 8$ and absence of the magnetic phase transition down to 0.15 K reveal the putative QSL ground state, this is sharply contrast to the observation of long-range AFM order in $YbCl_3$[22] and $SrYb_2O_4$ compounds.[48]

## ■ CONCLUSIONS

A new family of perfect honeycomb-lattice magnets $Ba_9RE_2(SiO_4)_6$ (RE=Ho-Yb) were successfully synthesized and magnetically characterized, where magnetic honeycomb layers with RE ions are alternatively stacked in an "ABCABC"-type fashion along the c-axis. Structure analysis reveals that the free of chemical antisite occupancies between magnetic $RE^{3+}$ ions and nonmagnetic $Ba^{2+}/Si^{4+}$ cations in these compounds, being a clear system to study the intrinsic honeycomb-lattice magnetism. All serial $Ba_9RE_2(SiO_4)_6$ (RE=Ho-Yb) polycrystals show the dominant antiferromagnetic interactions and absence of long-range magnetic order down to 2 K. The combination of isothermal magnetization and the ESR results suggest the different magnetic anisotropy for compounds containing different $RE^{3+}$ ions. More importantly, the synthesized $Ba_9Yb_2(SiO_4)_6$ single crystals don't show any long-range magnetic ordering down to 0.15 K, making it a promising Kitaev QSL material.

## ■ Notes

The authors declare no competing financial interest.

## ■ ACKNOWLEDGEMENTS

This work was supported by the National Natural Science Foundation of China (Grant No. 11874158), the Fundamental Research Funds of Guangdong Province (Grant No. 2022A1515010658) and Guangdong Basic and Applied Basic Research Foundation (Grant No.2022B1515120020). A portion of this work was supported by the synergetic extreme condition user facility (SECUF), and a portion of magnetic measurement was performed on the Steady High Magnetic Field Facilities, High Magnetic Field Laboratory. We would like to thank Shun Wang for his assistance on the specific heat measurement and thank the staff of the analysis center of Huazhong University of Science and Technology for their assistance in structural characterizations.

■ **REFERENCES**

■ Figure captions

**Figure 1.** (a) The schematic crystal structure of Ba$_9$Yb$_2$(SiO$_4$)$_6$, Yb atoms in orange, Si atoms in dark yellow, and three types of Ba atoms in light green, violet and blue-green, respectively. (b) The coordination environments of Yb, Ba1, Ba2, Ba3 and Si with oxygen atoms. (c) The "ABCABC"-type stacking fashion of magnetic Yb$^{3+}$ honeycomb layers along the *c*-axis. (d) The arrangement of magnetic Yb$^{3+}$ honeycomb lattice in the *ab* plane.

**Figure 2**. (a) The room-temperature powder X-ray diffraction (XRD) patterns of Ba$_9$RE$_2$(SiO$_4$)$_6$ (RE=Ho-Yb): the black circles denote experimental data, red lines show calculated patterns, the blue and green lines indicate the difference and background, and orange marks represent the positions of Bragg reflections. (b) The variation of lattice parameters as a function of RE ionic radii of Ba$_9$RE$_2$(SiO$_4$)$_6$. (c) The photograph of a representative Ba$_9$Yb$_2$(SiO$_4$)$_6$ single crystal.

**Figure 3**. (a) Temperature dependence of magnetic susceptibility $\chi$(T) and inverse susceptibility $1/\chi$(T) measured under $\mu_0 H$ = 0.5 T, the solid black lines show the Curie-Weiss fitting. (b) the isothermal $M(\mu_0 H)$ curves, the inset shows the $M(\mu_0 H)$ curves measured under pulsed magnetic field up to 42 T. (c) the ESR spectra at selected temperatures and (d) the temperature-dependent *g*-factors of Ba$_9$Ho$_2$(SiO$_4$)$_6$ polycrystals.

**Figure 4**. (a) Temperature dependence of magnetic susceptibility $\chi$(T) and inverse susceptibility $1/\chi$(T) measured under $\mu_0 H$ = 0.5 T, the solid black lines show the Curie-Weiss fitting. (b) the isothermal $M(\mu_0 H)$ curves, the dashed lines show the linear extrapolation on high-field magnetization ($\mu_0 H$ > 5.5 T) of $T$ = 2 K. (c) the ESR spectra at selected temperatures and (d) the temperature-dependent *g*-factors of Ba$_9$HoEr$_2$(SiO$_4$)$_6$ polycrystals.

**Figure 5.** (a) Temperature dependence of magnetic susceptibility $\chi$(T) and inverse susceptibility $1/\chi$(T) measured under $\mu_0 H$ = 0.5 T, the solid black lines show the Curie-Weiss fitting. (b) the isothermal $M(\mu_0 H)$ curves, the inset shows the $M(\mu_0 H)$ curves measured under pulsed magnetic

field up to 42 T. (c) the ESR spectra at selected temperatures and (d) the temperature-dependent $g$-factors of $Ba_9HoTm_2(SiO_4)_6$ polycrystals.

**Figure 6**. (a) Temperature dependence of magnetic susceptibility $\chi(T)$ and inverse susceptibility $1/\chi(T)$ measured in an applied field of 0.5 T, the solid black lines show the Curie-Weiss fitting. (b) the isothermal field-dependent magnetization $M(\mu_0H)$ curves, the dashed lines show the linear extrapolation on high-field magnetization ($\mu_0H > 5.5$ T) of $T = 2$ K. (c) the X-band ESR spectra at different temperatures and (d) the experimental and fitted ESR spectra of polycrystalline $Ba_9Yb_2(SiO_4)_6$ at $T = 2$ K, the red line represent Lorentzian fit.

**Figure 7**. (a) Temperature dependent magnetic susceptibility and inverse susceptibility $1/\chi(T)$ measured under $\mu_0H = 0.5$ T of $Ba_9Yb_2(SiO_4)_6$ single crystal for field along the $c$-axis and $ab$-plane. (b) the isothermal field-dependent magnetization $M(\mu_0H)$ curves of $Ba_9Yb_2(SiO_4)_6$ single crystal at 2 K for field along the $c$-axis and $ab$-plane, the data of $Ba_9Yb_2(SiO_4)_6$ powders is also present for comparison.

**Figure 8.** The Zero-field specific heat $C_p(T)$ of a $Ba_9Yb_2(SiO_4)_6$ single crystal, the inset shows the change of magnetic entropy $\Delta S_m(T)$ at low temperatures.

**Table 1**. The bond distances, bond angles and RE−RE distances of polycrystalline $Ba_9RE_2(SiO_4)_6$ (RE = Ho−Yb) samples.

**Table 2**. Crystal data and structure refinements of $Ba_9Yb_2(SiO_4)_6$ single crystals and its comparison with the data of polycrystals.

**Table 3**. The Curie-Weiss temperatures ($\theta_{CW}$) and effective magnetic moments ($\mu_{eff}$) determined from the Curie-Weiss fitting of magnetic susceptibility $\chi(T)$ for $Ba_9RE_2(SiO_4)_6$ (RE = Ho−Yb) compounds, the effective moment ($\mu_{fi}$) of free ions is calculated by $g[J(J + 1)]^{1/2}$.

Table 1. The bond distances, bond angles and RE−RE distances of polycrystalline $Ba_9RE_2(SiO_4)_6$ (RE = Ho−Yb) samples.

| RE | Ho | Er | Tm | Yb |
|---|---|---|---|---|
| Crystal system | Trigonal | Trigonal | Trigonal | Trigonal |
| Space group | R -3 | R -3 | R -3 | R -3 |
| RE−O2 (Å) | 2.19053(3) | 2.19015(3) | 2.18796(4) | 2.1804(3) |
| RE−O3 (Å) | 2.25237(3) | 2.25068(3) | 2.24990(4) | 2.2422(3) |
| Intraplane RE-RE (Å) | 5.79415(9) | 5.79128(9) | 5.78486(11) | 5.7644 |
| Interplane RE−RE (Å) | 7.31543(17) | 7.31150(17) | 7.3032(2) | 7.2791 |
| O2-RE−O2 (°) | 89.94 | 89.924 | 89.89 | 89.881 |
| O2-RE−O3 (°) | 95.027 | 95.043 | 95.077 | 95.085 |
| O3-RE−O3 (°) | 84.865 | 84.847 | 84.811 | 84.802 |
| Ba1-O3 (Å) | 2.84607(5) | 2.84499(5) | 2.84379(5) | 2.8342(3) |
| Ba2-O1 (Å) | 2.90903(5) | 2.90751(5) | 2.90438(6) | 2.8941(5) |
| Ba2-O2 (Å) | 3.11934(5) | 3.11851(5) | 3.11704(6) | 3.1065(4) |
| Ba2-O4 (Å) | 2.77560(4) | 2.77434(4) | 2.77297(5) | 2.7635(3) |
| Si-O1 (Å) | 1.58621(4) | 1.58648(4) | 1.58558(4) | 1.5803(3) |
| Si-O2 (Å) | 1.67001(3) | 1.66989(2) | 1.66789(3) | 1.66210(18) |
| Si-O3 (Å) | 1.65413(3) | 1.65402(3) | 1.65177(3) | 1.6460(2) |
| Si-O4 (Å) | 1.65413(3) | 1.63642(0) | 1.63322(2) | 1.62748(16) |

Table 2. Crystal data and structure refinements of $Ba_9Yb_2(SiO_4)_6$ single crystals and its comparison with data of polycrystals.

| Sample | Single crystal | Poly-crystal |
|---|---|---|
| Empirical formula | $Ba_9Yb_2(SiO_4)_6$ | |
| Formula weight | 2134.68 | |
| Crystal system | Trigonal | |
| Space group; Z | R -3; 3 | |
| a, b (Å) | 9.9969(7) | 9.993(3) |
| c (Å) | 22.1089(15) | 22.103(2) |
| α, β (°) | 90 | 90 |
| γ (°) | 120 | 120 |

| | |
|---|---|
| V (Å³) | 1913.5(3)   1910.065 |
| $\rho$ (cal) (g/cm³) | 5.557 |
| $\lambda$ (Å) | 0.71073 |
| $\theta$ (deg) | 2.527-32.188 |
| $\mu$ (mm⁻¹) | 21.274 |
| T (K) | 296(1) |
| F (000) | 2760 |
| Crystal size (mm³) | 0.03 × 0.02 × 0.02 |
| $R_p$ | 1.62 |
| $R_{wp}$ | 3.71 |
| Goodness of fit | 1.117 |

Table 3. The Curie-Weiss temperatures ($\theta_{CW}$) and effective magnetic moments ($\mu_{eff}$) determined from the Curie-Weiss fitting of magnetic susceptibility $\chi(T)$ for $Ba_9RE_2(SiO_4)_6$ (RE = Ho−Yb) compounds, the effective moment ($\mu_{fi}$) of free ions is calculated by $g[J(J + 1)]^{1/2}$.

| RE | High T fit | $\theta_{CW}$ (K) | $\mu_{eff}$ ($\mu_B$) | Low T fit | $\theta_{CW}$ (K) | $\mu_{eff}$ ($\mu_B$) | $\mu_{fi}$ ($\mu_B$) |
|---|---|---|---|---|---|---|---|
| Ho | 120 -300 K | - 8.89 | 10.90 | 40-70 K | - 3.50 | 10.70 | 10.60 |
| Er | 120 -300 K | - 7.53 | 9.93 | 5-15 K | - 2.43 | 9.14 | 9.59 |
| Tm | 120 -300 K | - 23.25 | 7.90 | 40-70 K | - 44.58 | 8.66 | 7.57 |
| Yb | 120 -300 K | - 112 | 4.86 | 5-15 K | - 1.18 | 2.62 | 4.54 |

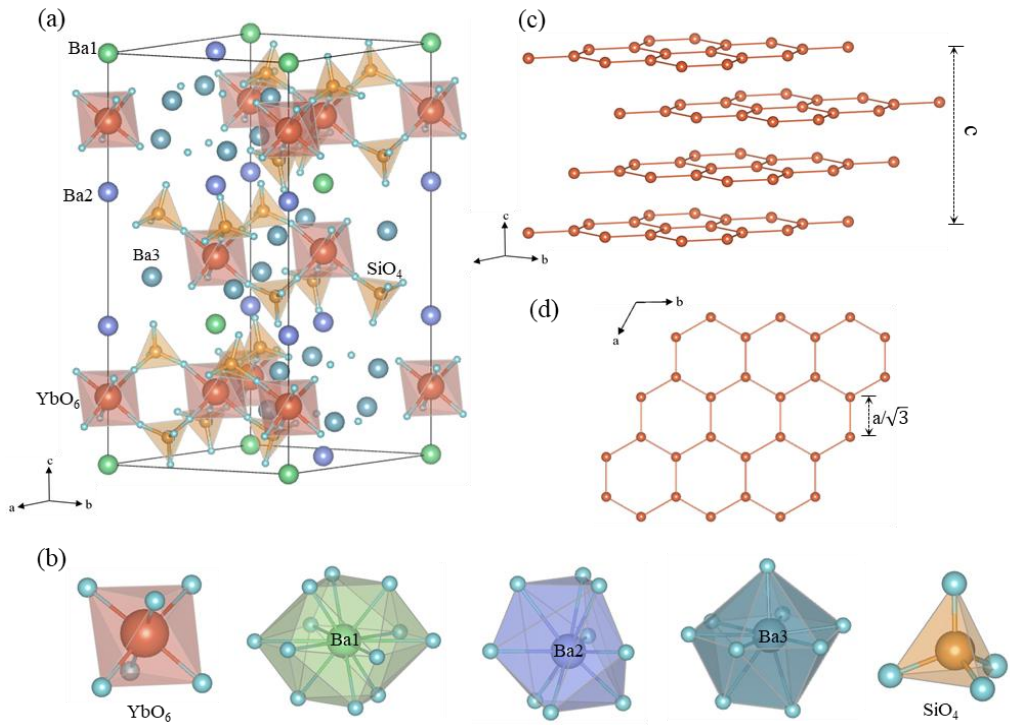

Figure 1

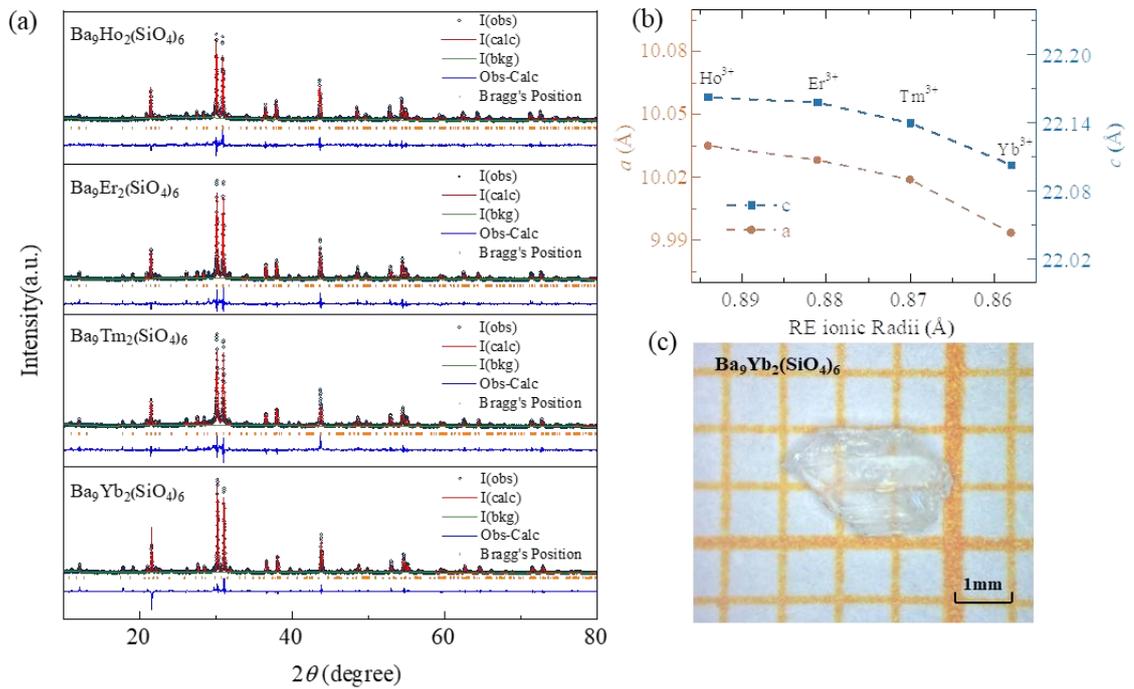

Figure 2

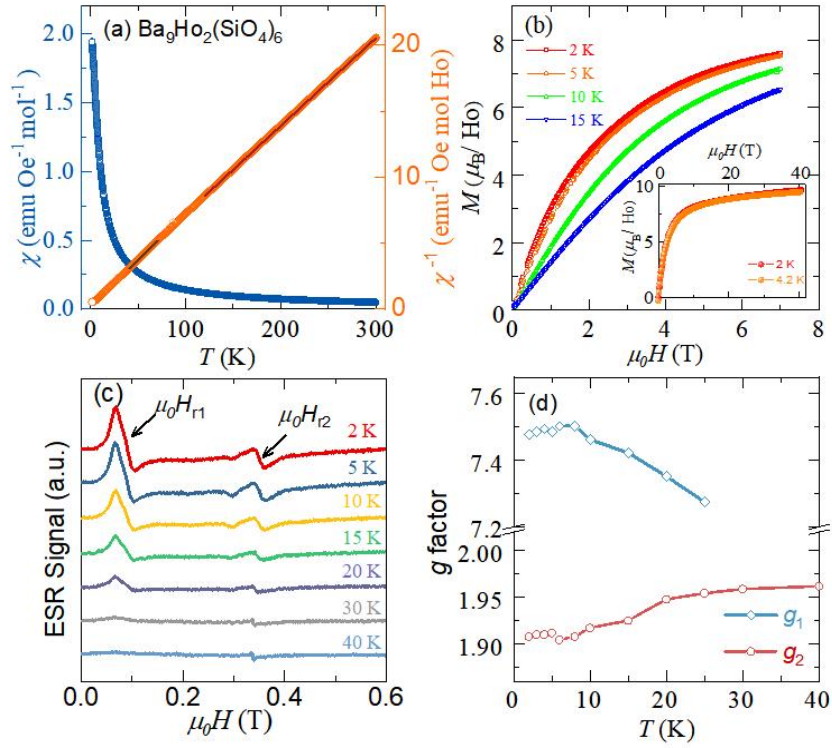

Figure 3

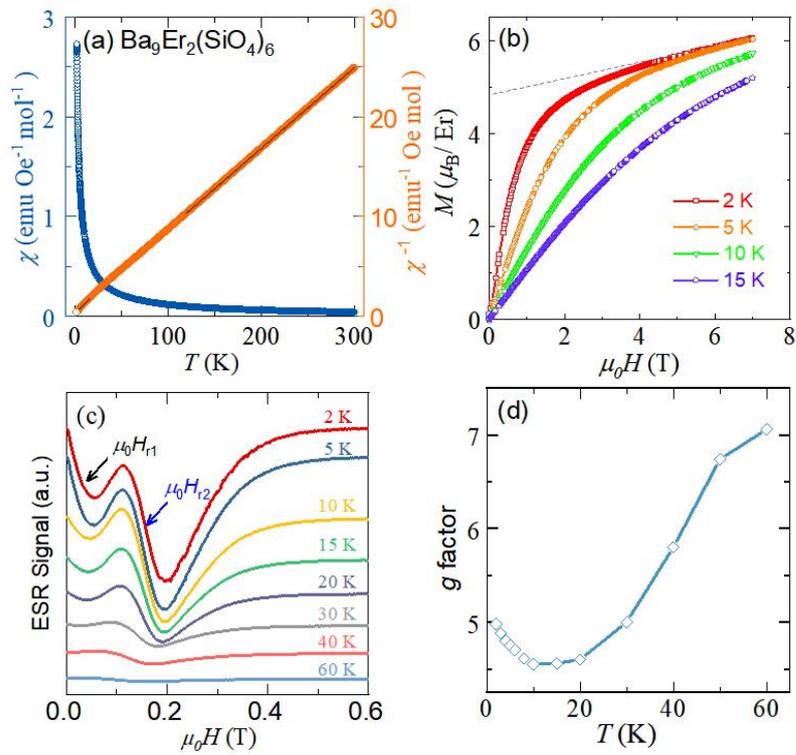

Figure 4

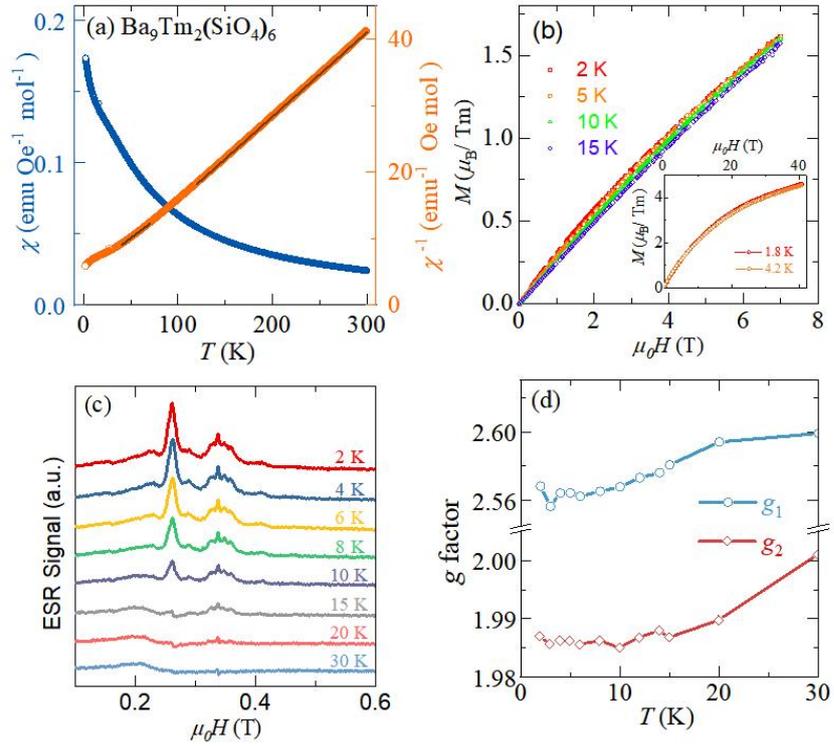

Figure 5

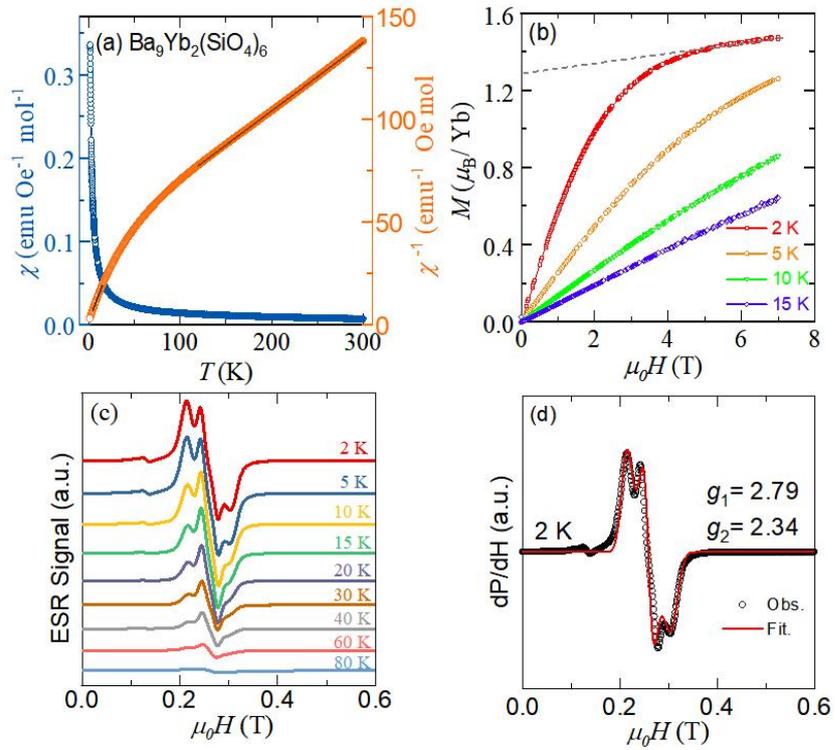

Figure 6

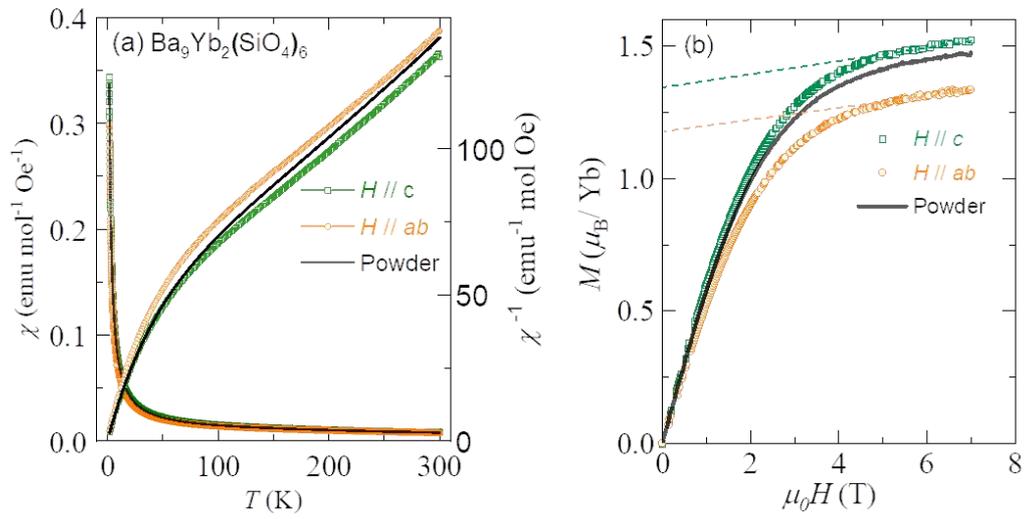

Figure 7

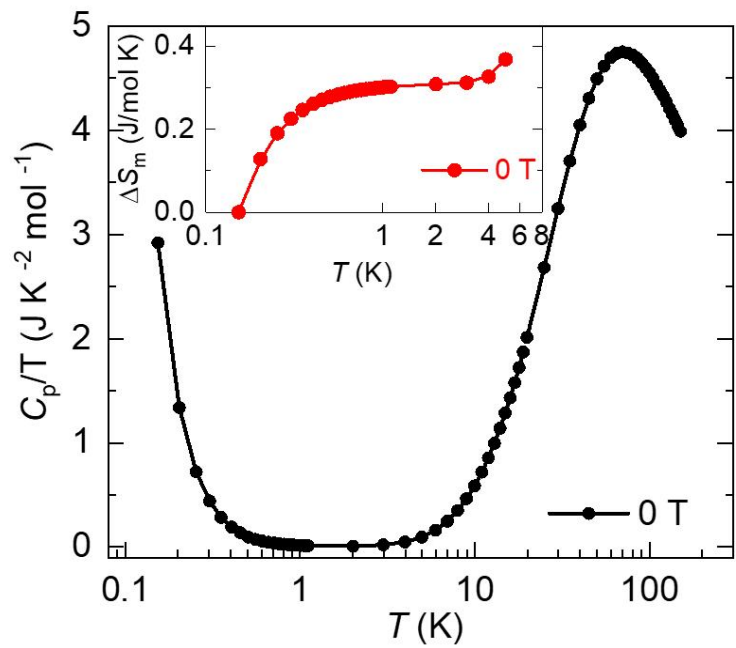

Figure 8